\documentclass[10pt,
superscriptaddress,
showpacs,preprintnumbers,
 amsmath,amssymb,
 prb,
floatfix,
lengthcheck,%
]{revtex4-1}
\usepackage{graphicx}
\usepackage{epstopdf} 
\usepackage{color}
\usepackage{dcolumn}
\usepackage{bm}
\usepackage{natbib}
\begin{document}
\title{Stability and energetics of 2D surface crystals \\
in liquid AuSi thin films and nanoscale droplets}
\author{Hailong Wang}
\affiliation{Department of Modern Mechanics, University of Science and Technology of China, Hefei 230027, China}
\author{Alireza Shahabi}
\affiliation{Group for Simulation and Theory of Atomic-Scale Material Phenomena (stAMP), Department of Mechanical and Industrial Engineering, Northeastern University, Boston, Massachusetts 02115, USA}
\author{Alain Karma}
\affiliation{Department of Physics, Northeastern University, Boston, Massachusetts 02115, USA}
\author{Moneesh Upmanyu}
\email{mupmanyu@neu.edu}
\affiliation{Group for Simulation and Theory of Atomic-Scale Material Phenomena (stAMP), Department of Mechanical and Industrial Engineering, Northeastern University, Boston, Massachusetts 02115, USA}

\begin{abstract}
Segregation at surfaces of metal-covalent binary liquids is often non-classical and in extreme cases such as AuSi, the surface crystallizes above the melting point. In this study, we employ atomic-scale computational frameworks 
to study the surface crystallization of AuSi films and droplets as a function of composition, temperature and size. For temperatures in the range $T_s^\ast=765-780$K above the melting point $(T_s^\ast\approx1.3\,T_m)$, both thin film and droplet surfaces undergo a first order transition, from a 2D Au$_2$Si crystalline phase to a laterally disordered yet stratified layer. The thin film surfaces exhibit an effective surface tension that increases with temperature and decreases with Si concentration. On the other hand, for droplets in the size range $10-30$\,nm, the bulk Laplace pressure alters the surface segregation as it occurs with respect to a strained bulk. Above $T_s^\ast$ the size effect on the surface tension is small, while for $T<T_s^\ast$ the surface layer is strained and composed of 2D crystallites separated by extended grain boundary scars that lead to large fluctuations in its energetics.
As a specific application, all-atom simulations of AuSi droplets on Si(111) substrate subject to Si surface flux show that the supersaturation dependent surface tension destabilizes the contact line {\it via} formation of a precursor wetting film on the solid-vapor interface, and has ramifications for size selection during VLS-based routes for nanowire growth. Our study sheds light on the interplay between stability and energetics of surfaces in these unique class of binary alloys and offers pathways for exploiting their surface structure for varied applications such as catalytic nanocrystal growth, dealloying, and polymer crystallization.     
\end{abstract}

\maketitle

\section{\label{sec:level1}Introduction}

Metal-covalent alloys form a versatile class of multifunctional material systems as their properties vary fundamentally with composition, from metallic alloys and amorphous glasses to crystalline semiconductors. These alloy systems often form low melting eutectics that serve as nanoscale catalysts for the growth of low-dimensional crystalline nanostructures such as semiconducting nanowires, inorganic nanotubes and nanoribbons. Examples of such nanostructure-catalyst systems include Si/Ge-Au, Si-Al, GaN/GaAs/GaP-Au/Ni, ZnO-Au, C/Co, C/Fe~\cite{nw:WagnerEllis:1964, nt:Iijima:2002, nw:PanWang:2001}. Several of these systems also serve as precursors for dealloying based routes for synthesis of nanoporous films and particles for catalysis and energy storage~\cite{dealloy:WadaKato:2016, dealloy:MaxwellBalk:2017, dealloy:ReinhardtHampp:2015}. The synthesis of these nanocrystals requires catalytic breakdown of the precursor gases on their surfaces at temperatures that usually exceed their melting points. The chemistry, structure and morphology of the liquid surfaces has a decisive effect on crystal nucleation and growth, and therefore directly impacts the quality of the as-grown nanostructures.

Past thin film studies have shown that the surfaces of several binary liquids are non-classical~\cite{tsf:YangRice:2007, tsf:ShpyrkoPershan:2006, tsf:ShpyrkoPershan:2007}. The surface is stratified, and in some cases, segregation triggers the formation of a solid-like surface layer that is stable above the bulk melting point. In the Au-Si system employed for the growth of Si nanowires (SiNWs), experimental studies on thin films have shown that the surface is Si-rich relative to the bulk~\cite{nw:FerralisMaboudian:2008, nw:LeeHwang:2010}. At near-eutectic temperatures, it is ordered preferentially through the thickness and capped with a laterally ordered crystalline layer~\cite{tsf:ShpyrkoPershan:2006, tsf:ShpyrkoPershan:2007}. A similar ordered surface structure was observed in  computations on nanoscale AuSi droplets~\cite{nw:WangUpmanyu:2013} and in experiments~\cite{nw:RossTersoff:2019}, suggesting that this phenomenon of surface crystallization is resilient to large surface curvatures. 

The surface structure and chemistry has obvious implications for catalysis as they both affect the chemical breakdown of the precursor gases on thin films and droplets. They also modify the surface energetics and this is an equally important aspect for the stability of supported droplets; the effective surface tension $\gamma_{lv}$ is a key ingredient for the Young's balance at the enveloping contact line formed by the intersection of two additional interfaces - substrate/crystal solid-vapor $\gamma_{sv}$, and substrate/crystal-particle solid-liquid $\gamma_{sl}$). During VLS growth of SiNWs for example, the resultant droplet morphology sets the surface chemical potential of the growing species, which in part controls the dynamics of the supersaturation that develops in the droplets, and therefore impacts the size and orientation selection of the nanocrystals~\cite{nw:RoperVoorhees:2007, nw:Mohammad:2008, nw:SchwarzTersoff:2009}. The surface energetics of nanoscale droplets is clearly crucial for controlled and scalable synthesis of these nanostructures {\it via} liquid mediated routes, yet the influencing factors remain largely unknown.  

In this study, we employ atomic-scale computational approaches to explore the interplay between surface structure and surface tension $\gamma_{lv}$ in crystalline surfaces of AuSi thin films and nanoscale droplets. The material system is motivated by the fact that AuSi nanoparticles continue to be important for scalable growth of SiNWs due to the combination of a low melting eutectic point [$X_{Si(E)}=19\%, T_E=636$\,K] and negligible solubility below the eutectic  temperature. More generally, it is a representative material system for fundamental studies on the vapor-liquid-solid (VLS) route for growth of semiconducting nanowires and related nanostructures, for crystal nucleation and growth at the nanoscale~\cite{nw:GivargizovChernov:1973, nw:Givargizov:1975, nw:KodambakaRoss:2006, nw:KimRoss:2008, nw:DavidGentile:2008, nw:KimRoss:2009, nw:WenTersoffRoss:2011, nw:HemesathVoorheesLauhon:2012}, surface catalysis, and for stability and dynamics of templated alloying and dealloying~\cite{dealloy:ReinhardtHampp:2015}. 

The article is organized as follows: we first  quantify the liquid surface energetics at and near-equilibrium by analyzing the surface structure and extracting the surface tension as a function of the growth temperature and droplet composition, $\gamma_{lv}\equiv \gamma_{lv}(X_{Si}, T)$. The computations are performed for both thin films and droplets with varying radii to capture the size effect, $\gamma_{lv}\equiv \gamma_{lv}(R)$. As self-consistent validation of the results, we perform simulations of AuSi droplets on Si substrates and compare their stability at equilibrium, and under near-equilibrium conditions  by subjecting the droplet surfaces to a steady-state Si flux. We conclude with a discussion of the results as they relate to VLS growth of semiconducting nanowires, and more general catalytic growth of crystalline nanofilaments.

\section{Computational Methods}
The atomic-scale computations are based on a recently developed AuSi empirical interatomic potential~\cite{nw:DongareZhigilei:2009}. The classical approximation is crucial as it allows access to length- and time-scales necessary for extraction of surface structures and surface tension. The potential accurately predicts the stability of AuSi clusters and droplets and is tailored to reproduce the binary Au-Si phase diagram~\cite{nw:WangUpmanyu:2013}. In particular, it recovers the signature low melting eutectic, with a eutectic point that is in good agreement with experimental values, i.e. $T_E=590$\,K, $X_{Si}^E=33\%$.


The classical approach employed to extract the structure and energetics of liquid surfaces is based on equilibrium molecular dynamics (MD) simulations of near-eutectic AuSi alloys. The simulations are performed for thin film and nanoscale droplet geometries, and for temperatures and compositions in the range $T=590$-$1100$\,K and $X_{Si}=20-45\%$, respectively. For both classes of simulation geometries, the initial configuration consists of a randomly mixed alloy with prescribed composition $X_{Si}$ and mean inter-atomic distances corresponding to the liquid alloy density. Semi-grandcanonical Monte-Carlo (SGMC) simulations followed by MD simulations are employed to obtain a fully equilibrated alloy~\cite{surfseg:Foiles:1985}. The SGMC equilibration is modified to preserve the stoichiometry via a combination of translational and exchange moves  (ratio fixed at 1:10). Typically, this extends to tens of million translation steps. To further equilibrate the structure at the desired temperature, the SGMC output is subject to additional canonical MD simulations in excess of a nanosecond using a N\"ose-Hoover thermostat~\cite{book:AllenTildesley:1989} with a fixed time step of $1$\,fs. This sequence of simulations is performed until the interaction energy and surface density profiles converge. 

Alternatively, in all of the cases, we have performed the equilibration directly with MD, and the results reliably converge to those obtained by SGMC simulations in less than a nanosecond. Data runs performed following equilibration consist of equilibrium MD simulations for an average of $10$\,ns, and the configurations are stored every 10 time steps for structural analyses. 

\subsection{Capillary Fluctuation Method, CFM}
The surface tension of the thin films is extracted from the power spectrum of the surface capillary waves, based on the capillary fluctuation method (CFM)~\cite{fec:BuffStillinger:1965, fec:Weeks:1977}. The fluctuations in profiles of a free liquid-vapor (vacuum) surface is monitored and time-averaged over uncorrelated atomic configurations. In the limit of small slopes, decomposition in Fourier space yields a mode dependent fluctuation amplitude,
\begin{equation}
\label{eq:CFM}
\langle|{A(k)}|^2\rangle=\frac{k_BT}{A_s k^2} \;\;\frac{1}{\gamma+\gamma^{\prime\prime}}\,,
\end{equation}
where 
$A_s=bl$ is the projected surface area and $\gamma+\gamma^{\prime\prime}$ is the surface stiffness. For liquid surfaces with inclination independent energetics, the stiffness is the surface tension of the planar film, $\gamma+\gamma^{\prime\prime}=\gamma_{lv}^0$.

\begin{figure*}[htb]
\includegraphics[width=1.75\columnwidth]{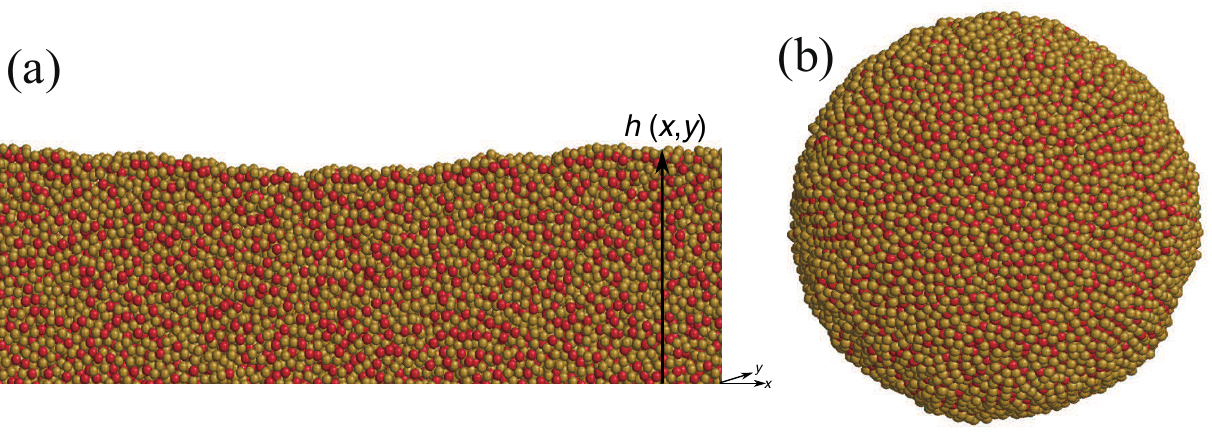}
\caption{Atomic configurations of the simulation cell geometries used for extracting  $X_{Si}=0.33\%$ and $T=873$\,K. (a) Cross-section of a ribbon-like liquid slab showing morphological fluctuations of one of the two surfaces. Cubic liquid slabs are used for extracting the surface stresses (not shown). (b) Equilibrated isolated liquid droplets used for extracting size-dependent surface tension. Here and elsewhere, red and gold colors denote Si and Au atoms, respectively.\label{fig:simCell}}
\end{figure*}

Simulations of liquid slabs with ribbon-like geometries each of length $l=50$\,nm and thickness $b\approx2$\,nm are used to extract the fluctuation spectrum. Periodic boundary conditions are applied in-plane ($x-y$) and two $z$-surfaces are held free. Film thicknesses of $h\approx25$\,nm ensure that the two free surfaces do not interact. In total, the size of each simulation is $\approx80000$ atoms. As an example, Fig.~\ref{fig:simCell}a shows one of the two fluctuating surfaces with $X_{Si}=33\%$  at $T=873\,^\circ$K.  The surface profile $h(x, y)$ is extracted by monitoring the positions of the surface atoms and the surface structure is characterized by the local density within slices of thickness 0.05\,{\rm \AA}.  

\subsection{Surface Stress Method, SSM}
An alternate approach is based on the relation between the surface tension and the stress state of liquid surfaces~\cite{intseg:YuStroud:1997, intpot:WebbGrest:1986}. The mechanical definition takes the form    
\begin{equation}
\gamma_{lv}^0=\frac{1}{2}(\sigma_{xx}^s+\sigma_{yy}^s-2\sigma_{zz}^s),
\end{equation}
where $\sigma_{\alpha\beta}$ is surface stress tensor, $(\alpha, \beta)\equiv (x, y, z)$. Its components are calculated from the atomic interactions and the trajectories of the surface atoms~\cite{def:VitekEgami:1987},
\begin{equation}
\sigma_{\alpha\beta}^s=-\frac{1}{S}\sum_i[mv_i^{\alpha}v_i^{\beta}+\frac{1}{4}\sum_j(f_{ij}^{\alpha}r_{ij}^{\beta}+f_{ij}^{\beta}r_{ij}^{\alpha})]
\end{equation}
where $S$ is the effective surface area, $m_i$ and $v_i$ are the atomic mass and velocity, and $f_{ij}$ is the force exerted by each $j^{th}$ neighbor located at a distance $r_{ij}$ within the cut-off distance associated with the short-range interaction potential. 
Unlike the ribbon-like slab geometry used for CFM simulations, fluctuations in surface morphology are undesirable as they lead to spatial variations in the surface stresses. The fluctuations decay with the surface aspect ratio~\cite{sold:Karma:1993} and we therefore employ $8.5\,{\rm nm}\times8.5\,{\rm nm}\times8.5\,{\rm nm}$ cubic slabs for this class of simulations. The densities and surface stresses are monitored in surface slices of thickness 0.05\,{\rm\AA} within equilibrium MD simulations.

\subsection{Laplace Pressure Method, LPM}
The size dependence $\gamma_{lv} (R)$ is studied by performing similar computations on isolated nanoscopic droplets. For a droplet in equilibrium with its vapor, the (bulk) Laplace pressure that balances the surface capillary forces is related to the surface tension by the well-known Gibbs-Thomson equation~\cite{book:RowlinsonWidom:1982, surf:Tolman:1948},
\begin{equation}
\label{eq:GibbsThomson}
{\Delta}P=\frac{2\gamma_{lv}^0}{R_e}(1-\frac{\delta}{R_e}+...)
\end{equation}
where $\gamma_{lv}^0$ is the surface tension of a planar surface and $R_e$ is the surface within which the chemical excess $\Gamma$ is zero, that is 
\begin{equation}
\label{eq:EqRadius}
\Gamma=\int_0^\infty[\rho(r)-\rho_b(r)] r^2dr=0,
\end{equation}
with $\rho_b=\rho_l$ for $r<R_e$ and $\rho_b=\rho_v$ for $r>R_e$. Here, $\rho_l$ and $\rho_v$ are the densities of the bulk liquid and the vapor, respectively. This study is limited to droplet-vacuum surface and therefore we take $\rho_v=0$. 
Equation~\ref{eq:GibbsThomson} yields the size dependent surface tension of the droplet,
\begin{equation}
\label{eq:Tolman}
\gamma_{lv}=\gamma_{lv}^0(1-\frac{2\delta}{R_e}+...).
\end{equation}
The extent of the size dependence is embodied by the length scale $\delta$, also known as the Tolman length~\cite{int:Tolman:1949}. 
 

The pressure differential across the droplet surface is calculated using the virial stress within the bulk,
\begin{equation}
\Delta P=-\frac{1}{3}(\sigma_{xx}+\sigma_{yy}+\sigma_{zz}),
\end{equation}
where the stresses $\sigma_{\alpha\beta}$ are components of the atomistic virial stress tensor~\cite{def:VitekEgami:1987},
\begin{align}
\sigma_{\alpha\beta}=\frac{1}{V}\sum_i \Big[ mv_i^{\alpha}v_i^{\beta}
+\frac{1}{4}\sum_j(f_{ij}^{\alpha}r_{ij}^{\beta}+f_{ij}^{\beta}r_{ij}^{\alpha}) \Big]\,,\nonumber
\end{align}
extracted and averaged over a bulk volume $V=4{\pi}R^3/3$ with $R\le R_e$.
\begin{figure}[bp]
\includegraphics[width=\columnwidth]{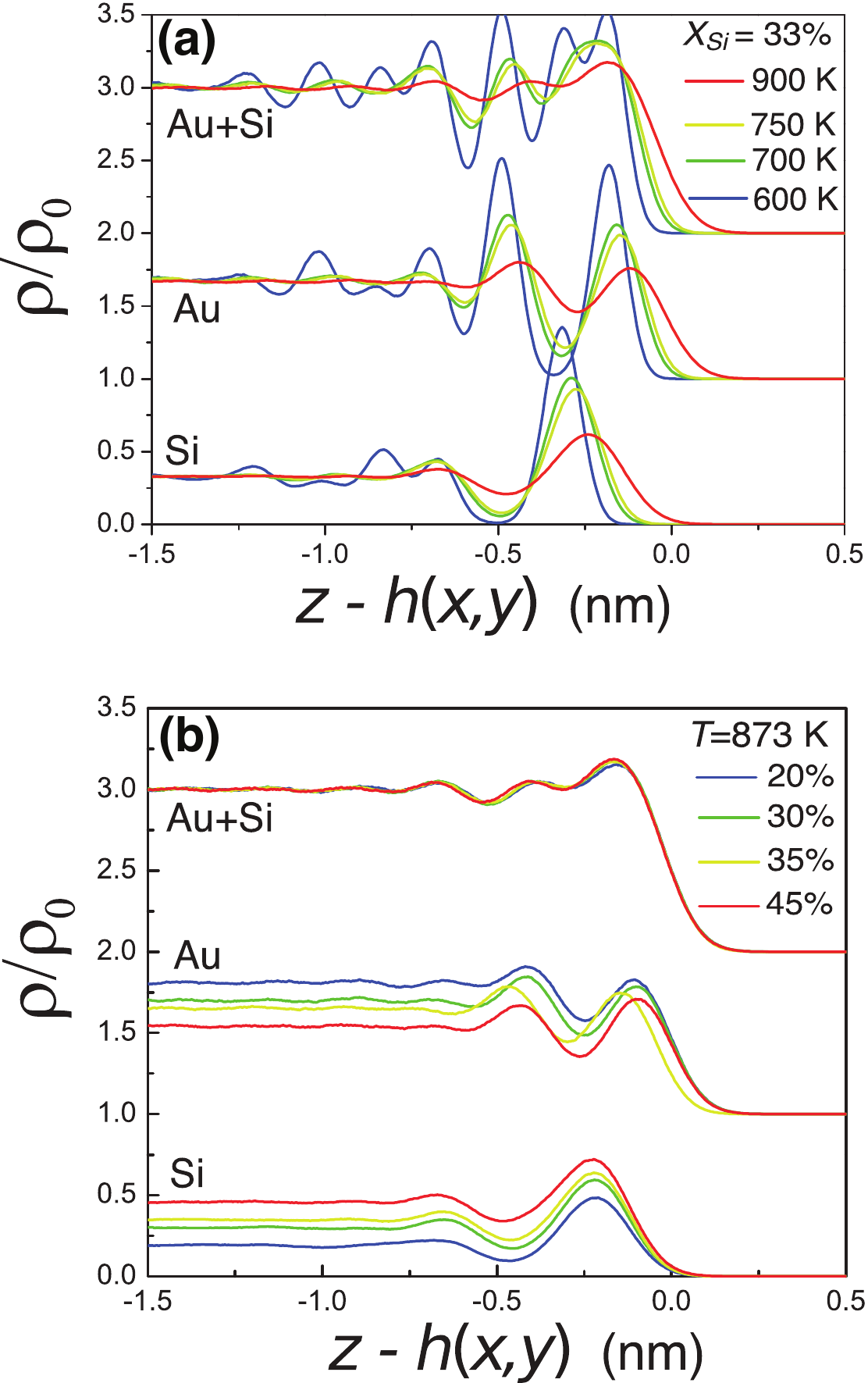}
\caption{Through thickness variations of density profiles $\rho_{Au+Si}(z)$, $\rho_{Au}(z)$ and $\rho_{Si}(z)$ in liquid AuSi slabs extracted using equilibrium MD simulations, (a) for varying temperatures at at $X_{Si}=33\%$, and (b) varying compositions at $T=873$\,K. The profiles are calculated with respect to the local height profile $h(x,y)$ of the film and averaged over the two free surfaces. The component plots are shifted along the vertical axis for clarity. The densities are normalized to the atom number densities of bulk alloys at the corresponding composition and temperature. \label{fig:densityFilms}}
\end{figure}

The density and Laplace pressure are recorded in shells with thickness $0.1$\,nm every 10 time steps. The simulations are performed for droplets with varying sizes in the range $2R=10-30$\,nm ($32,000-256,000$ atoms). In each case, the equimolar droplet radius $R_e$ is based on the definition  Eq.~\ref{eq:EqRadius}. 
Comparisons with the results from the thin film computations yield a direct measure of the size effect.

\section{Results}
\subsection{Thin Films}
\begin{figure*}[htp]
\includegraphics[width=1.9\columnwidth]{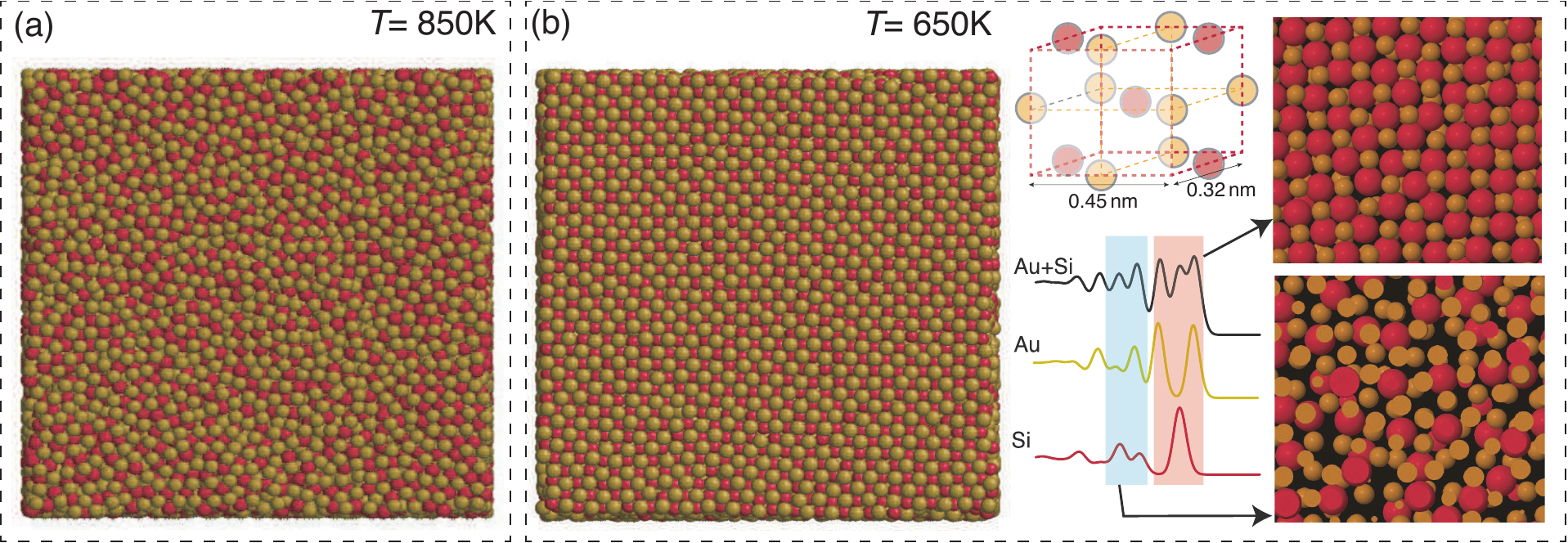}
\caption{Atomic configurations of the surface structure observed in thin film computations at a) $850$\,K and b) $650$\,K. The capping Au monolayer is disordered in a) and exhibits lateral order as the temperature is lowered in b). (bottom middle panel, b) The overall and component density profiles within $0.32$\,nm thick surface and subsurface layers shaded light blue and red, respectively. The corresponding atomic configurations of one of the two surfaces is shown in the right panel. (top middle panel, b) The Au$_2$Si unit cell associated with the lateral ordering of the surface layer. See text for details.\label{fig:AuOrdering}}
\end{figure*}

Surface segregation is evident in the sectional view of the liquid slab shown in Fig.~\ref{fig:simCell}a. We quantify it by extracting the through-thickness variation of the normalized density $\rho(z)/\bar{\rho}$ of both Au and Si. Figure~\ref{fig:densityFilms}a shows the variation averaged over the two thin films surfaces at the eutectic composition $X_{Si}=0.33\%$ and at varying temperatures above the bulk eutectic melting point, $T_m=T_E$. In all instances, the surface is on average denser compared to the bulk, consistent with recent {\it ab-initio} and classical MD simulations~\cite{nw:LeeHwang:2010, nw:WangUpmanyu:2013}. Closer examination of the density profiles near the surface reveals peaks in Au and Si densities. At higher temperatures $T>800$\,K, the component density profiles indicate sub-surface Si segregation relative to the bulk that alternates with smaller Au peaks, characteristic of a stratified surface layer. 
Figure~\ref{fig:AuOrdering}a shows the top view of the surface at $T=850$\,K. Lateral order along the film surface is absent. In all instances, the surface is capped by a Au sub-monolayer, consistent with the sustained catalysis of Si-precursors on these surfaces as well as the stability of Au adatoms on Si-rich surfaces~\cite{surf:Zhang:2001}.   
\begin{figure}[bp]
\includegraphics[width=\columnwidth]{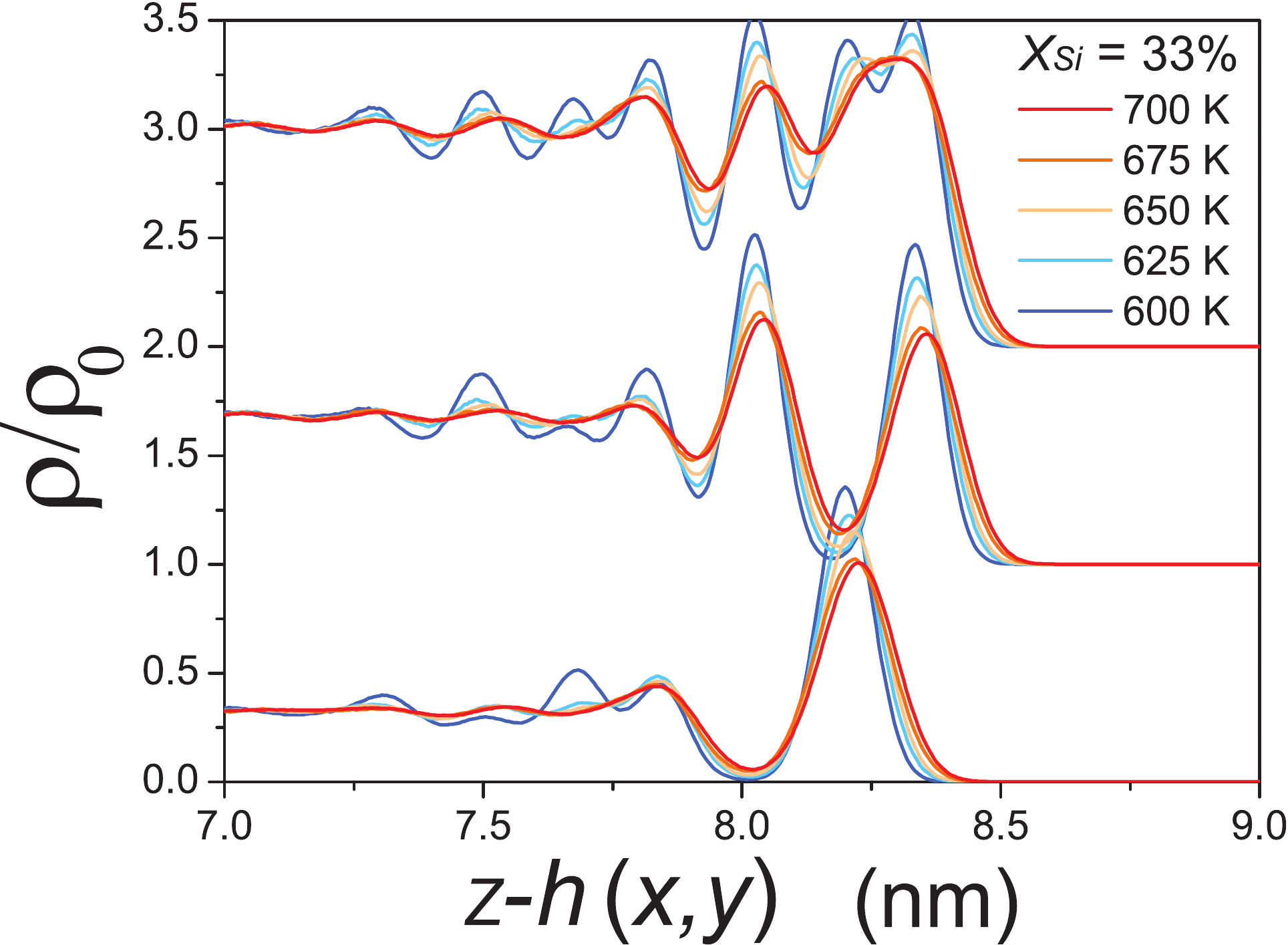}
\caption{Same as in Fig.~\ref{fig:densityFilms}a but for a narrower temperature range, $T=600-700$\,K. \label{fig:densityFilms600-700}}
\end{figure}

Lowering the temperature has a dramatic effect on the density profiles. The Si and Au surface peaks increase in their intensity. Consequently, the surface excess and the through-thickness order are both enhanced. The extent over which the peaks decay into the bulk also increases, signifying an increase in the overall depth of the surface stratification. Below $T\approx750$\,K, lateral crystallization along the surface occurs. The capping Au layer transitions to an ordered monolayer. The long-range order extends to the Si-rich sub-surface based on the surface peaks in the overall density profiles and also visible in the top view of one of the free surfaces at $T=650$\,K (Fig.~\ref{fig:AuOrdering}b). The density profiles reveal that the crystallization enhances through-thickness density of the surface layer. For example, we see the emergence of a bilayer peak in the density profile composed of the Si layer and the capping Au layer.  

The thickness of the crystallized layer is $0.32\pm0.02$\,nm and does not vary with temperature. Careful examination of surface slices reveals a 2D square Si lattice (point group $p4m$) with a lattice parameter $a=0.32$\,nm, sandwiched by two identically structured Au-lattices shifted by one-half the lattice parameter. Taken {\it in toto}, the three layers order into an A(Au)-B(Si)-A(Au) stacking sequence (top right, Fig.~\ref{fig:AuOrdering}b); a schematic of the Au$_2$Si unit cell is also shown in the figure. The surface structure is unlike any bulk crystalline phase in this system. It is close to the Au$_3$Si$_2$ structure reported in {\it ab-initio} simulations of Au$_70$Si$_30$ liquid~\cite{nw:LeeHwang:2010} but differs from past reports on the structure low and high temperature AuSi surface phases extracted using grazing incidence x-ray diffraction (GIXD)~\cite{tsf:ShpyrkoPershan:2007, tsf:MechlerPershan:2010}, primarily due to the presence of the capping Au-monolayer. 

Immediately below the crystalline layer, the surface is stratified yet laterally disordered (bottom right, Fig.~\ref{fig:AuOrdering}b), and the disorder increases away from the surface (Fig.~\ref{fig:densityFilms}a). Unlike the crystallized layer, the thickness of the stratified layer  markedly increases as the temperature approaches the bulk melting point. To see this clearly, Fig.~\ref{fig:densityFilms600-700} shows in variation in the density profiles for a narrower range of temperatures, from $T=600-700$\,K. The component profiles show emergence of new peaks away from the surface within the liquid, indicative of stratification into Au- and Si-rich layers as the temperature decreases. We see a sharp increase in the extent of stratification below $650$\,K. The crystallized layer does not exhibit significant changes in its density, consistent with the stability of this low temperature (LT) surface phase~\cite{tsf:ShpyrkoPershan:2007, tsf:MechlerPershan:2010, nw:RossTersoff:2019}. 

Increasing $X_{Si}$ has a qualitatively different effect. Figure~\ref{fig:densityFilms}b shows the density profiles with varying composition at $T=873\,K$. The surface is stratified but laterally disordered indicating that the temperature is above the surface crystallization temperature for these range of compositions. The overall density profile as well as the width of the ordered surface layer do not change significantly, yet the surface becomes increasingly Si rich.  As the temperature is lowered below $T\approx750\,K$, we again see the appearance of laterally ordered surface layer. The thickness of the 2D crystal does not change significantly within the range of compositions explored here, indicating that the changes in the Si excess at the surface are confined to the subsurface stratified layers.

\begin{figure}[bp]
\includegraphics[width=0.9\columnwidth]{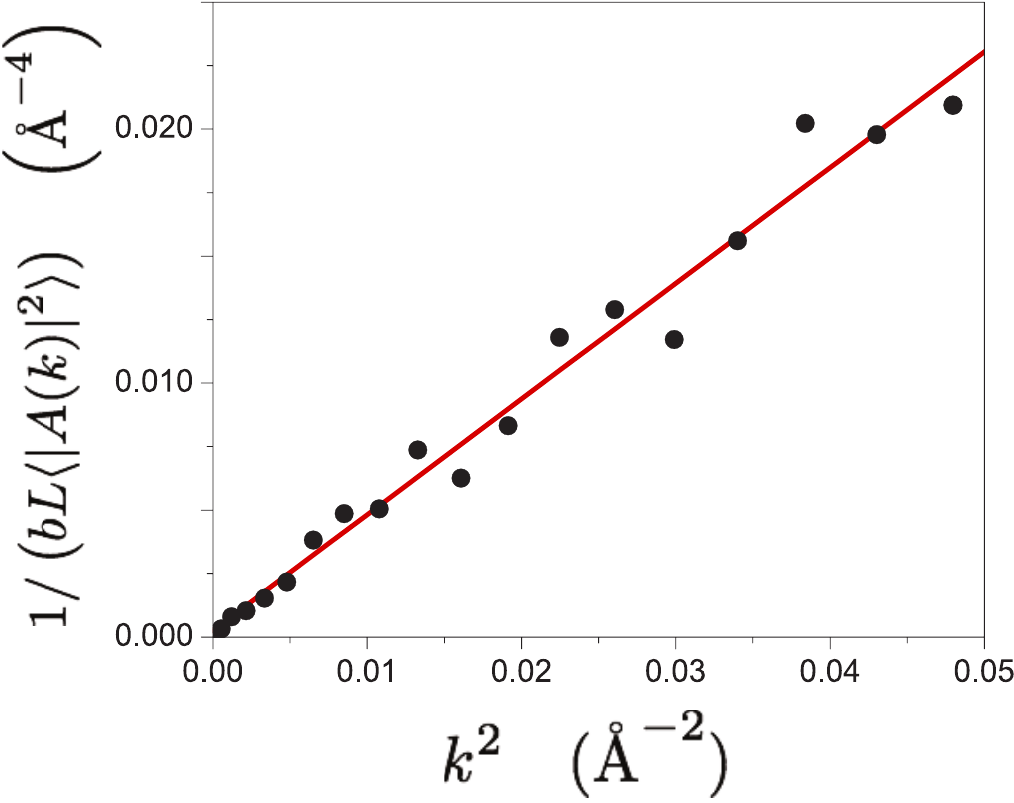}
 \caption{Ensemble averaged fluctuation spectrum $1/ \langle{{\mid}A(k){\mid}}^2\rangle bW$ vs $k^2$ of the two surfaces for the liquid slab shown in Fig.~\ref{fig:simCell}, for $X_{Si}=33\%$ and at $T=873$\,K. The slope of the linear fit yields the surface tension, $\gamma_{lv}^0=0.58\pm0.02$\,J/m$^2$. \label{fig:powerSpectrum}}
\end{figure}

We delegate a detailed structural characterization of the surface crystal to a later study, and extract the effective surface tension of the crystallized layer as a function of temperature and composition by monitoring the fluctuations and surface stresses in the liquid slabs. Figure~\ref{fig:powerSpectrum} shows the fluctuation spectrum at $X_{Si}=33\%$ and $T=873$\,K.  The surface fluctuations are statistically significant, in particular the long wavelength (small $k$) amplitudes extend to several interatomic distances over the nanosecond-scale simulations. The spectrum is representative of the behavior observed for all composition ranges at $T=873$\,K in that $(\langle|A(k)|^2\rangle A_s)^{-1}$ increases linearly with $k^2$, i.e. the mean square amplitude $\langle|A(k)|^2\rangle$ decays as $k^{-2}$ and the stratified surface behaves as a classical liquid surface~\cite{fec:Weeks:1977}. Following Eq.~\ref{eq:CFM}, the slope yields the thin film surface tension, $\gamma_{lv}^0=0.58\pm0.02$\,J/m$^2$.

\begin{figure}[thp]
\includegraphics[width=\columnwidth]{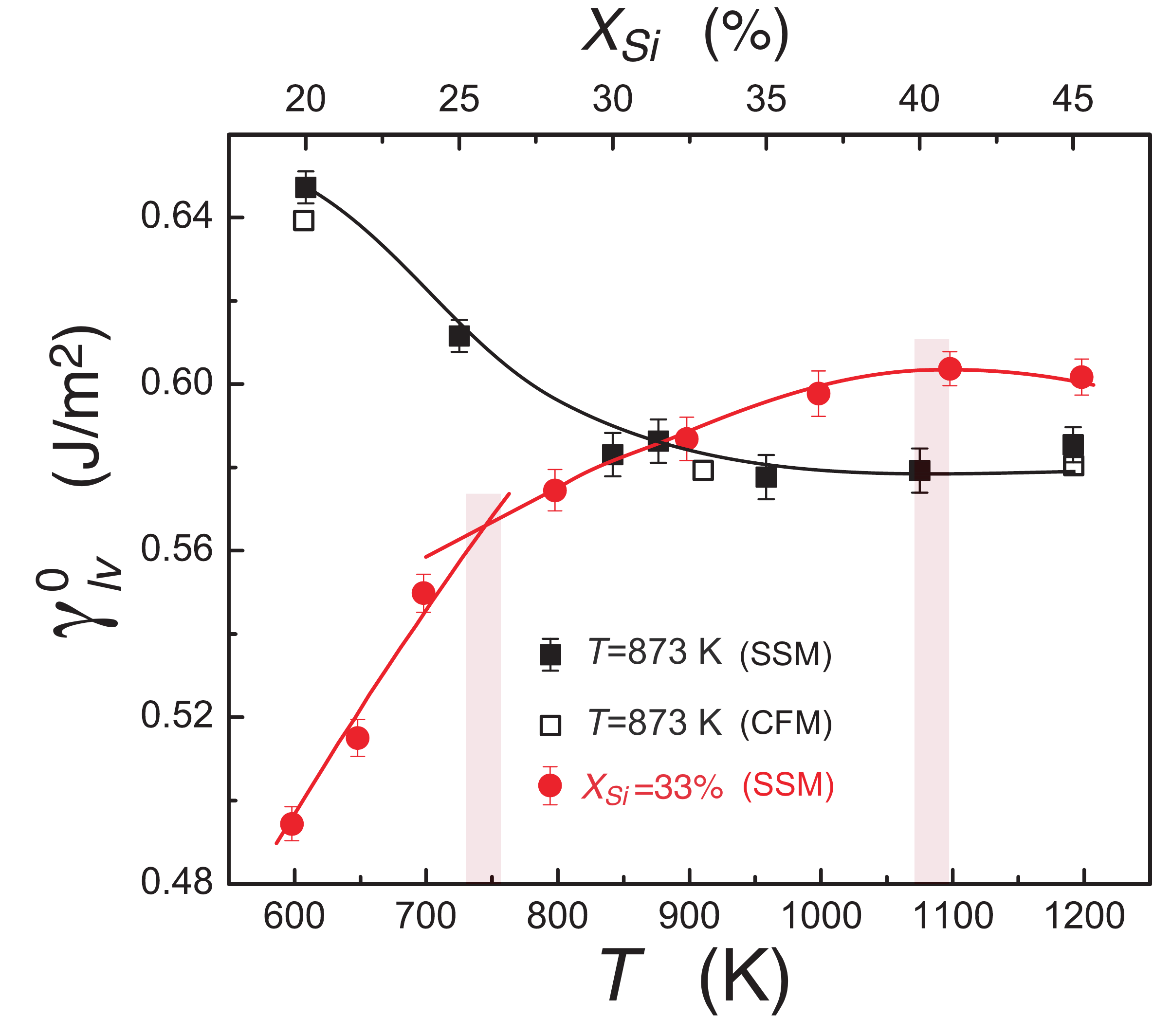}
 \caption{Temperature and composition dependence of the planar AuSi surface tension $\gamma_{lv}$. The composition dependence is extracted at $T=873$\,K using capillary fluctuation method (CFM, open black squares) and surface stress method (SSM, solid black squares). The temperature dependence is extracted at $X_{Si}=33\%$ using SSM (solid red circles). The two shaded temperature intervals represent distinct changes in the temperature dependence $d\gamma_{lv}/dT$. \label{fig:filmTension}}
\end{figure}

The variation in planar film surface tension with composition extracted using CFM is plotted in Fig.~\ref{fig:filmTension}. Increasing $X_{Si}$ lowers the surface tension and the sensitivity is higher for hypoeutectic alloys: $\gamma_{lv}^0=0.64$\,J/m$^2$ for $X_{Si}=20\%$ and it decreases to $\gamma_{lv}^0=0.58$\,J/m$^2$ for $X_{Si}=33\%$. Above the eutectic composition, the decrease is smaller and for $X_{Si}\ge42\%$, the trend is reversed as the surface tension exhibits a small increase with concentration. The minimum in the surface tension corresponds to $\gamma^0_{lv}\approx0.58$\,J/m$^2$ over the composition range $X_{Si}=33-40\%$.

The surface fluctuations of the surface become increasingly suppressed at lower temperatures. We therefore rely on SSM to extract the temperature dependence $\gamma_{lv}(T)$. As validation, the compositional variation extracted {\it via} SSM is in excellent agreement with the CFM results over the entire composition range (Fig.~\ref{fig:filmTension}). 
The temperature dependence at fixed composition $X_{Si}=33\%$ is plotted in Fig.~\ref{fig:filmTension}. At $T=600$\,K, just above the eutectic temperature ($T/T_m=1.02$), the surface tension is quite low, $\gamma_{lv}^0=0.49$\,J/m$^2$. It  {\it increases} with temperature and then saturates to a value of $\gamma_{lv}^0=0.61$\,J/m$^2$ above $\approx1050$\,K. The increase in the surface tension with temperature  (slope $d\gamma_{lv}/dT>0$) is indicative of surface prefreezing. Our results show that while the width of crystallized surface layer does not diverge as we approach $T=T_m$, the divergence in the thickness of the stratified layer of the adjoining liquid layers leads to a thermodynamics behavior not unlike the behavior of a prefrozen surface layer. 

The surface tension does not increase uniformly with temperature. Rather we observe two regimes with characteristic values of $d\gamma_{lv}/dT$ separated at $T\approx750$\,K. Since the slope is an indirect measure of the order within the layer, the transition signifies a change in the extent of crystalline order within the surface layer, quite possibly a slow order-disorder transition of the surface layer from fully crystallized surface layer to a laterally disordered yet stratified surface. 


\subsection{Nanoscale Droplets}
\begin{figure}[htp]
\includegraphics[width=0.9\columnwidth]{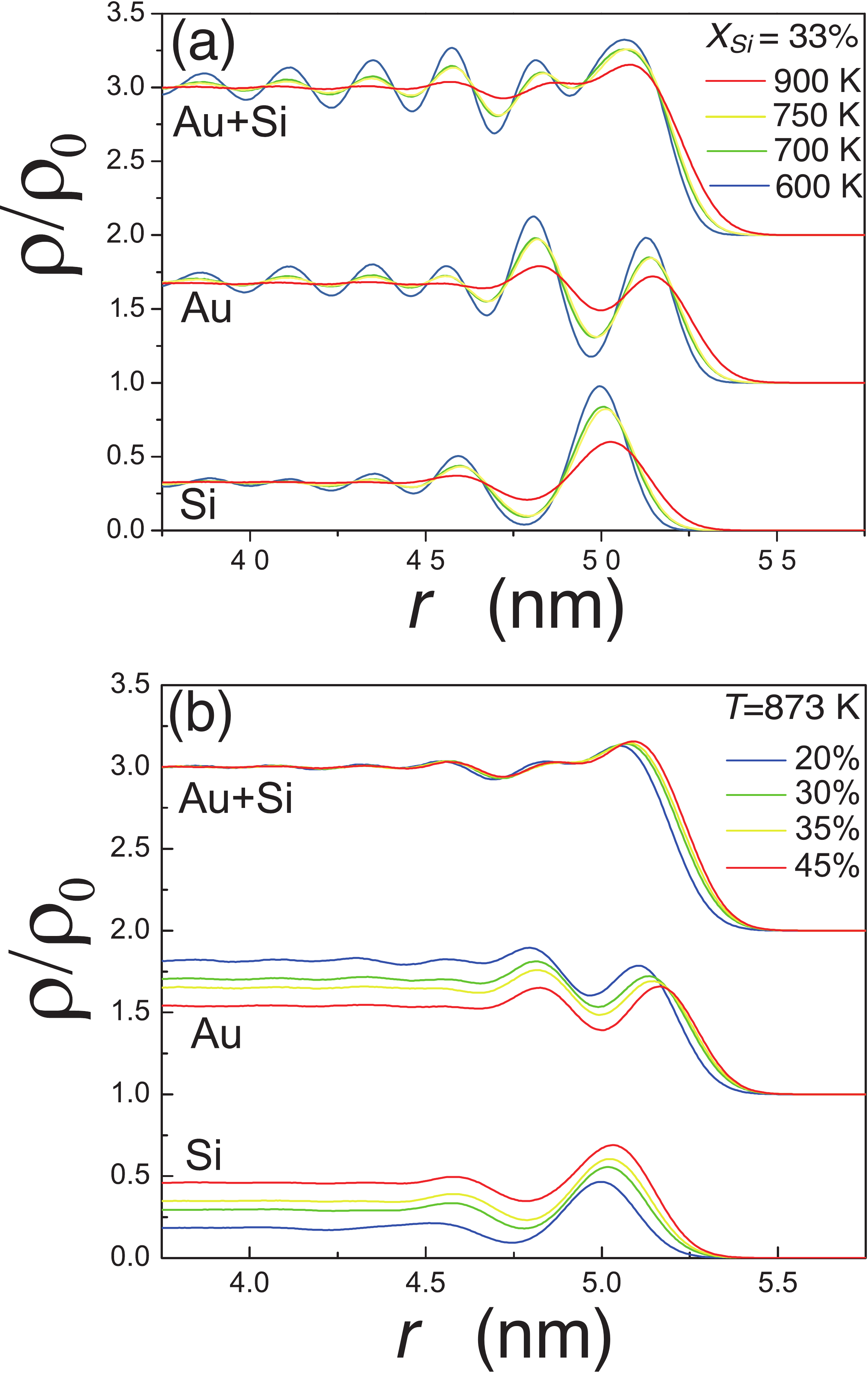}
\caption{Radial variation of the normalized density profiles $\rho(r)/\bar{\rho}$ for a $2R=10$\,nm AuSi droplet, for the same set of temperatures and compositions as in Fig.~\ref{fig:densityFilms}.\label{fig:densityDroplets}}
\end{figure}

Figure~\ref{fig:densityDroplets} shows the temperature and composition dependence of the radial density profile $\rho(r)/\bar{\rho}$ within a $2R=10$\,nm diameter AuSi droplet. The surface of the droplet consists of terminating Au submonolayer with alternating Si-rich and Au-rich subsurface layers, similar to the surface structure observed for thin films.  At $T=873$\,K, the subsurface is again silicon-rich and gradually decays to the bulk composition. The intensities of the Si and Au peaks and the thickness of the ordered surface layer both increase with decreasing temperature. 
\begin{figure*}[thp]
\includegraphics[width=2\columnwidth]{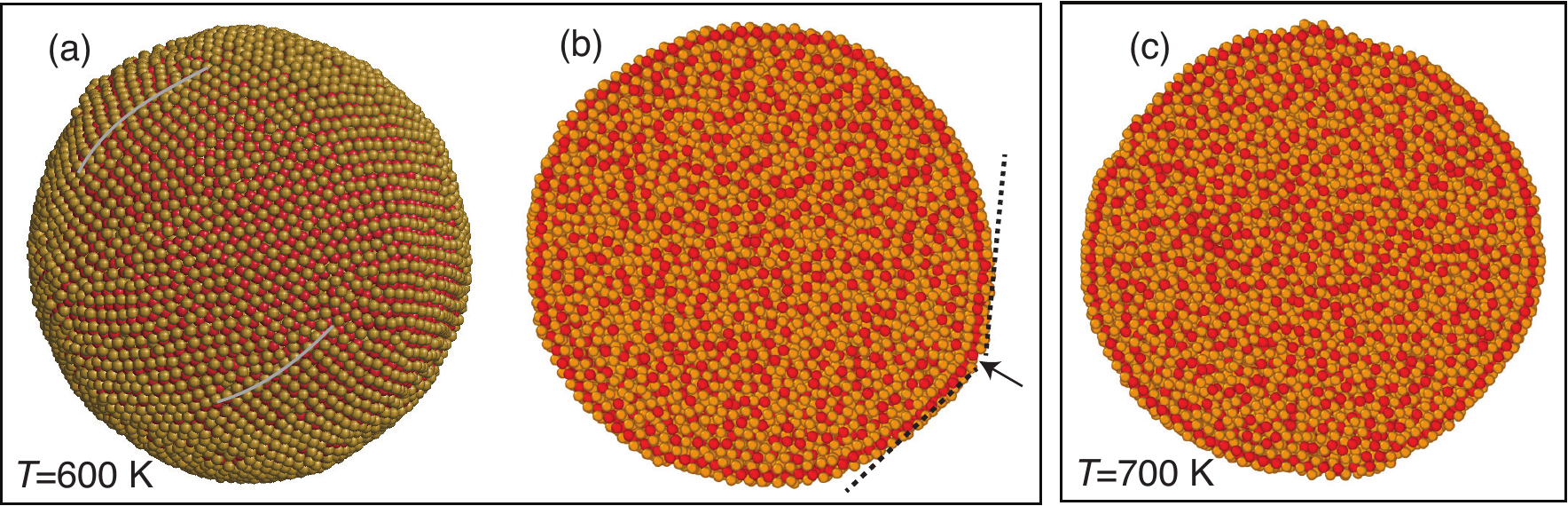}
\caption{(c) Atomic configuration of a $2R=10$\,nm droplet equilibrated at $T=600\,$K. The Au surface monolayer consists of ordered domains separated by finite length grain boundaries or grain boundary scars~\cite{2dlayers:BauschNelsonWeitz:2003}. Two such instances of these surface defects are indicated (solid gray lines). Note the segregation of surface Au atoms at these extended defects. (b) Midsection of the droplet that that shows the surface segregation and crystallization. (c) Same as in (a), but for $T=700$\,K.  \label{fig:AuOrderingDroplet}}
\end{figure*}

As the temperature is lowered closer to the eutectic temperature we see the appearance of long-range lateral order at the surface, visible in the equilibrium droplet configuration at $T=600$\,K  shown in Fig.~\ref{fig:AuOrderingDroplet}a. The corresponding midsection configuration (Fig.~\ref{fig:AuOrderingDroplet}b) reveals that the crystallization is limited to the surface with stratification in the adjoining liquid. The surface crystal exhibits some faceting separated by extended defects indicating that the surface tension is orientation dependent. Two such intersecting facets are indicated in the midsection. On increasing the temperature, the facets become less favorable, that is the surface tension becomes increasingly isotropic with respect to the surface orientation. As a direct comparison, Fig.~\ref{fig:AuOrderingDroplet}c shows the same midsection at  $T=700$\,K.

The overall trends in the behavior are qualitatively similar to that observed in thin films, there are some important deviations. Comparisons with density profiles for thin films at comparable temperatures (Fig.~\ref{fig:densityFilms}) show that the peak intensities associated with surface segregation are smaller for the droplets. Near $T_E$, the ordering within the surface layers is relatively suppressed. This is likely due to the large surface curvature of the droplet that is geometrically accommodated by point and extended defects within the crystallized surface layer (Fig.~\ref{fig:AuOrderingDroplet}). Specifically, we see the appearance of several surface grains separated by finite length quasi-2D grain boundaries, not unlike the geometrically necessary grain boundary scars observed within crystals on spherical surfaces~\cite{2dlayers:BauschNelsonWeitz:2003}. Furthermore, although the compositional variation is qualitatively unchanged in that the subsurface becomes Si-rich with increasing  $X_{Si}$ (Fig.~\ref{fig:densityDroplets}), unlike thin films the Au and Si peaks at the surface shift outward as their intensity changes, consistent with expansion of the ordered surface layer possibly driven by changes in the surface stress state relative to the thin films.

\begin{figure}[bp]
\includegraphics[width=0.9\columnwidth]{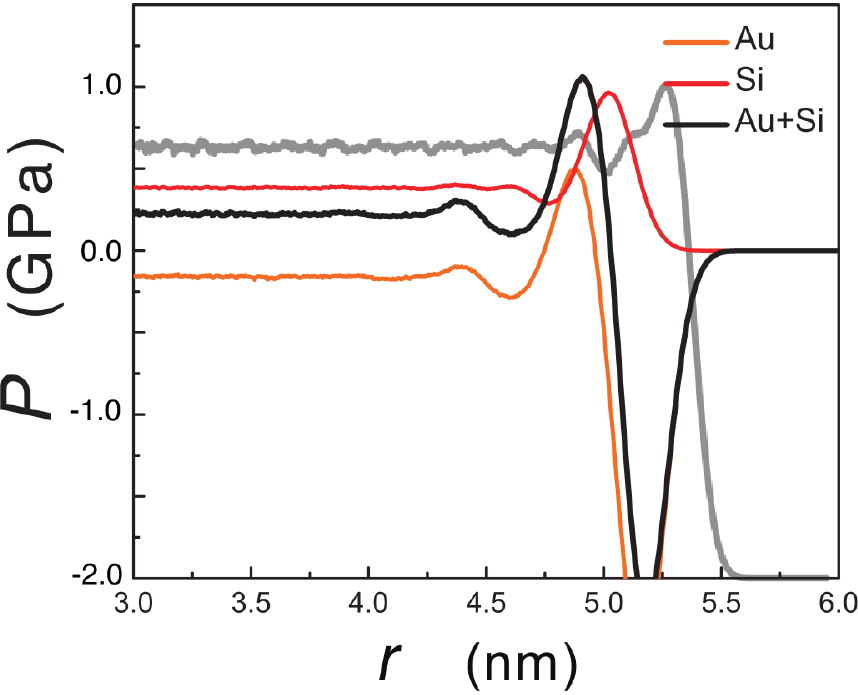}
 \caption{Radial distribution of the virial pressure $P_{Au+Si}(r)$, $P_{Au}(r)$ and $P_{Si}(r)$ within a $2R=10$\,nm diameter AuSi droplet at $X_{Si}=33\%$ and $873$\,K. The gray solid line is the corresponding overall density profile (scale not shown) that is superposed to help correlate the pressure distribution to the surface crystallization.
 \label{fig:LaplacePressure}}
\end{figure}

\begin{figure*}[htbp]
\includegraphics[width=1.8\columnwidth]{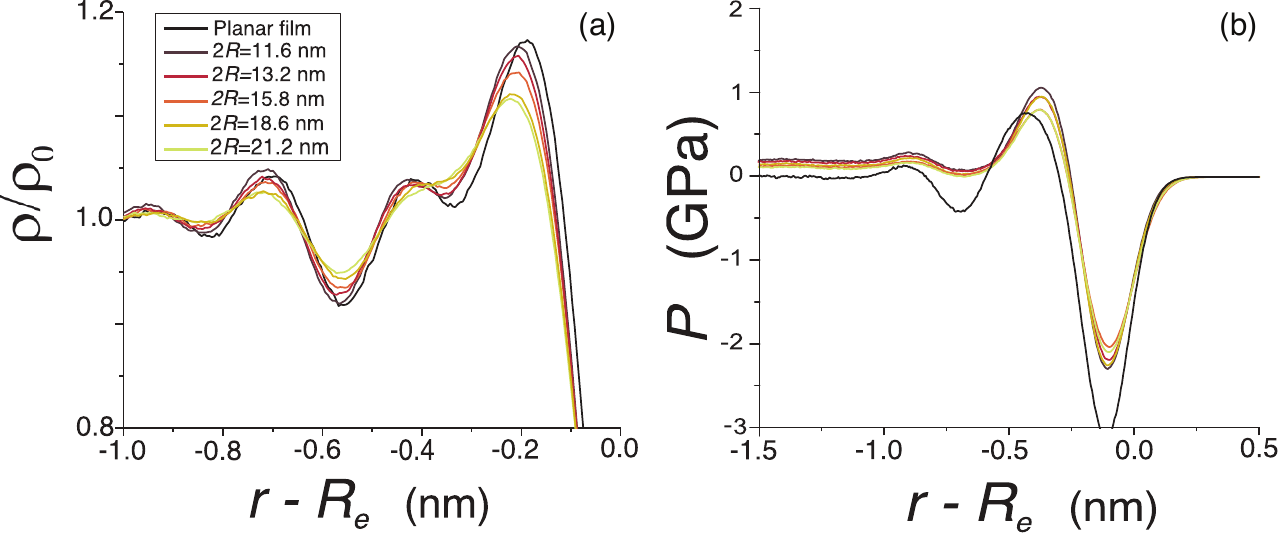}
 \caption{Size dependence of (left) normalized radial density variation and (right) pressure distribution in equilibrated AuSi thin film and droplets with varying radii in the range $2R=10-22$\,nm, with $X_{Si}=33\%$ and at $T=873\,$K. \label{fig:sizeVarDensityPressure873}}
\end{figure*}
To develop an understanding of these variations, we characterize the stress distribution within the droplets. Figure~\ref{fig:LaplacePressure} shows the radial distribution of the virial pressure within a droplet with $X_{Si}=33\%$ and at $T=873$\,K. To help correlate the pressure distribution with the segregation profile, the overall normalized density distribution is also shown in the plot. The Laplace pressure is constant and positive for $r<4.2$\,nm.  The pressure at the surface varies non-monotonically within a $1.2$\,nm thick layer. The ensemble averaged pressure difference across the entire surface layer is $\langle \Delta P\rangle\approx0.2$\,GPa. Examination of the component variations show that the Si-rich subsurface is compressive ($P>0$) while the capping Au monolayer results in tensile stresses ($P<0$). The non-monotonic variations in the pressure are strongly correlated with segregation at the surface. Examination of the component density profiles (not shown) confirms that in general Si excess is associated with compressive stresses while Au excess counteracts their build up. Accordingly, the subsurface stratification of Si is associated with compressive stresses while the capping Au monolayer is associated with a tensile stress. 

The pressure distribution and density profiles are similar for droplets in the size range studied here, $2R=10-30$\,nm. The size range is large enough such that the surface segregation does not modify the composition in the bulk~\cite{np:RingeMarks:2015}. Figure~\ref{fig:sizeVarDensityPressure873}a shows their radial variation at $X_{Si}=33\%$ and at $T=873\,$K. The distribution normal to the surface of the planar thin film is also plotted for comparison.  The segregation profile for the largest droplet simulated here ($2R=21.2$\,nm) deviates significantly from that for the planar thin film, suggesting that the effect of the surface curvature is significant. The reduced intensities of the surface peaks indicates that the large droplet surface curvature suppresses both Si and Au segregation to the surface, thereby reducing the through thickness stratification of the droplet surface. The size dependence shows that decreasing the droplet size results in an increase in the intensity of the surface and subsurface peaks as well as an increase in the equimolar radii, implying that the surface becomes more stratified with surface curvature for range of sizes simulated here. 

Decreasing the size at constant temperature and surface tension is analogous to increasing temperature at constant size as the Gibbs-Thomson effect causes a depression in the melting point of the droplets at smaller sizes, $\Delta T_m\propto \gamma_{lv}/R$. However, the trends in the simulations are consistent with {\it decreasing} temperature. Evidently, the enhanced stratification in smaller droplets is unrelated to size dependence of the droplet melting point.

Changes in the pressure distribution within the droplets with size show deviations from the variation observed in planar films (Figure~\ref{fig:sizeVarDensityPressure873}b). The compressive stresses associated with subsurface stratification of Si and Au increase and the tensile stresses within the capping monolayer in the droplets are considerably suppressed.
Notably, the bulk of the droplet is strained due to the Laplace pressure that balances the surface tension.  Careful examination of the size dependence of these extrema shows that their intensities increase with decreasing size, again indicating that the variation is not due to size dependence of the droplet melting point. Rather, these variations reflect the increasing Laplace pressure within the bulk of the droplet. Then, the surface (chemical) excess at the surface occur with respect to a strained bulk, and pressure effects on surface segregation become important. For a droplet so stressed, the leading order correction to the segregation-based surface excess for species $i$ is~\cite{gbseg:McLean:1957, gbseg:Lejcek:2010} 
\begin{equation}
\label{eq:segregationPressure}
\frac{\Gamma_i (p=P)}{\Gamma_i (p=0)} = \exp\left[-\frac{\langle P\rangle\Delta v_i}{RT}\right],
\end{equation}
where $\Delta v_i$ is activation volume of the segregating species at the surface and $\langle P\rangle$ is the average (external) pressure in the bulk. For a two component system such as the Au-Si system considered here, the activation volume is the relative change in the partial molar volumes of the two species between the surface and bulk. As an example, denoting $\Omega^S$ and $\Omega^B$ as the average partial molar volumes at the surface and bulk respectively, the change in activation volume for Au can be expressed as
\begin{equation}
\label{eq:actVolChange}
\Delta v_{Au}=(\Omega_{Au}^S-\Omega_{Au}^B)-(\Omega_{Si}^S-\Omega_{Si}^B).
\end{equation} 
\begin{figure}[bp]
\includegraphics[width=\columnwidth]{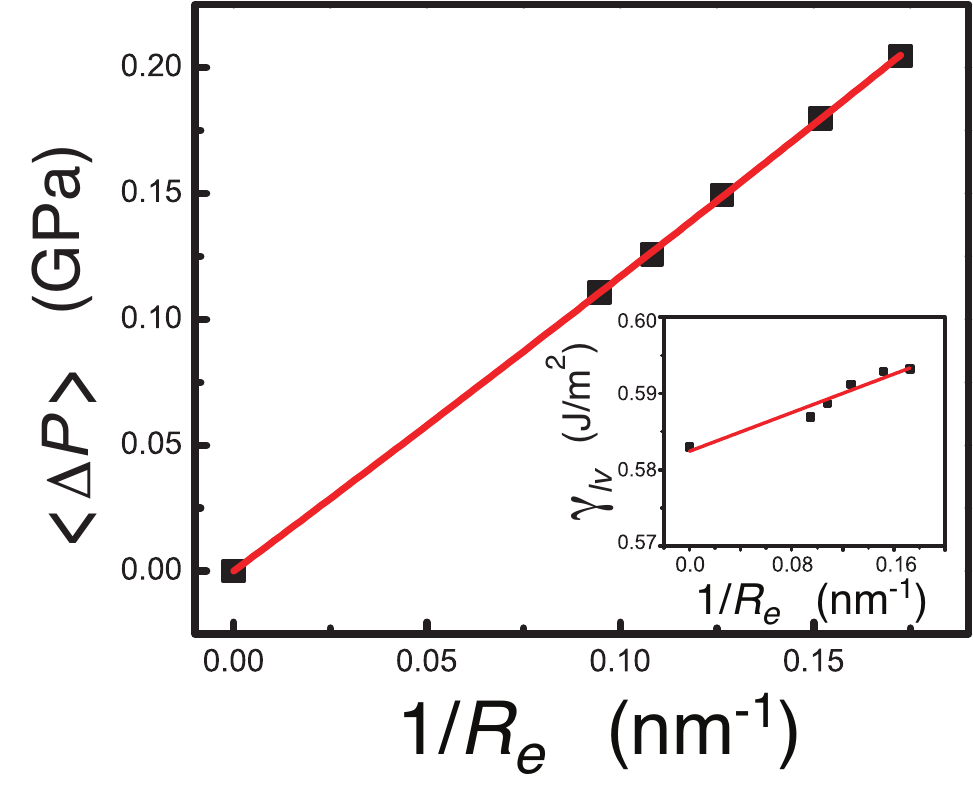}
 \caption{Plot of the ensemble average pressure differential across the surface $\langle \Delta P\rangle$ vs equimolar droplet curvature $\langle 1/R_e \rangle$ for droplets with sizes in the range $2R=10-30$\,nm. The solid red line is a second-order polynomial fit based on Eq.~\ref{eq:GibbsThomson}.
 (inset) Plot of the size dependence of the effective droplet surface tension $\gamma_{lv}=R_e/2\,\langle \Delta P\rangle$. \label{fig:dropletEnergy}}
\end{figure}

Following Eqs.~\ref{eq:segregationPressure} and~\ref{eq:actVolChange}, the thermodynamic interplay between the surface segregation and the pressure distribution in the droplet is mediated by the activation volume changes between the bulk and surface. The interplay offers a qualitative explanation for the extracted trends in the size dependence of the density and pressure distribution. For example, the increase in the Au monolayer segregation with decreasing size is a consequence of $\Delta v_{Au}>0$.
Then, the combination of the tensile stress in the layer and the increasingly strained bulk of the droplet results in increase in the Au excess. Similarly, the enhanced stratification of the subsurface due to Si segregation follows from the combined effect of $\Delta v_{Si}<0$ and the fact that Si excess is associated with a compressive stress. Then, the strained bulk of the droplet results in a decrease in Si excess in the subsurface layer with decreasing size, thereby enhancing the overall stratification of the surface layer. In the case of the planar film, the segregation is further enhanced yet the surface stress state quickly vanishes away from the bulk. As such, the activation volume changes have little effect on the surface chemical excesses.     

The size dependence of the average pressure within the droplet $\langle \Delta P\rangle$ allows an independent measure of the droplet surface tension. Figure~\ref{fig:LaplacePressure} shows the variation of the $\langle \Delta P\rangle$ with the droplet curvature $1/R_e$ extracted at $T=873$\,K and $X_{Si}=33\%$. The pressure differential increases almost linearly with the droplet curvature. A second order polynomial fit based on Eq.~\ref{eq:GibbsThomson} extrapolated to $R=\infty$ yields the planar surface tension, $\gamma_{lv}^0=0.582 \pm 0.01$\,J/m$^2$. The Tolman length associated with the size effect of the surface tension is negative large, $\delta=-0.5\pm0.1$\,nm. The planar surface tension is in excellent agreement with the value extracted from computations of thin films, $\gamma_{lv}^0=0.58$\,J/m$^2$. A negative value of the Tolman length implies a larger effective droplet radius $R_{eff}=R_e-\delta>R_e$, where $R_{eff}$ is the droplet radius that makes the Laplace equation exact. Its magnitude is of the order of the liquid interatomic distance, implying that the size effect is small. The surface tension associated with the effective radius, $\gamma_{lv}=R_{eff}/2\,\langle \Delta P\rangle$ increases non-linearly within the narrow range $0.586-0.594$\,J/m$^2$ with the surface curvature for droplet sizes in the range $2R=10-30$\,nm (inset, Fig.~\ref{fig:LaplacePressure}b). 
\begin{figure*}[htbp]
\includegraphics[width=1.8\columnwidth]{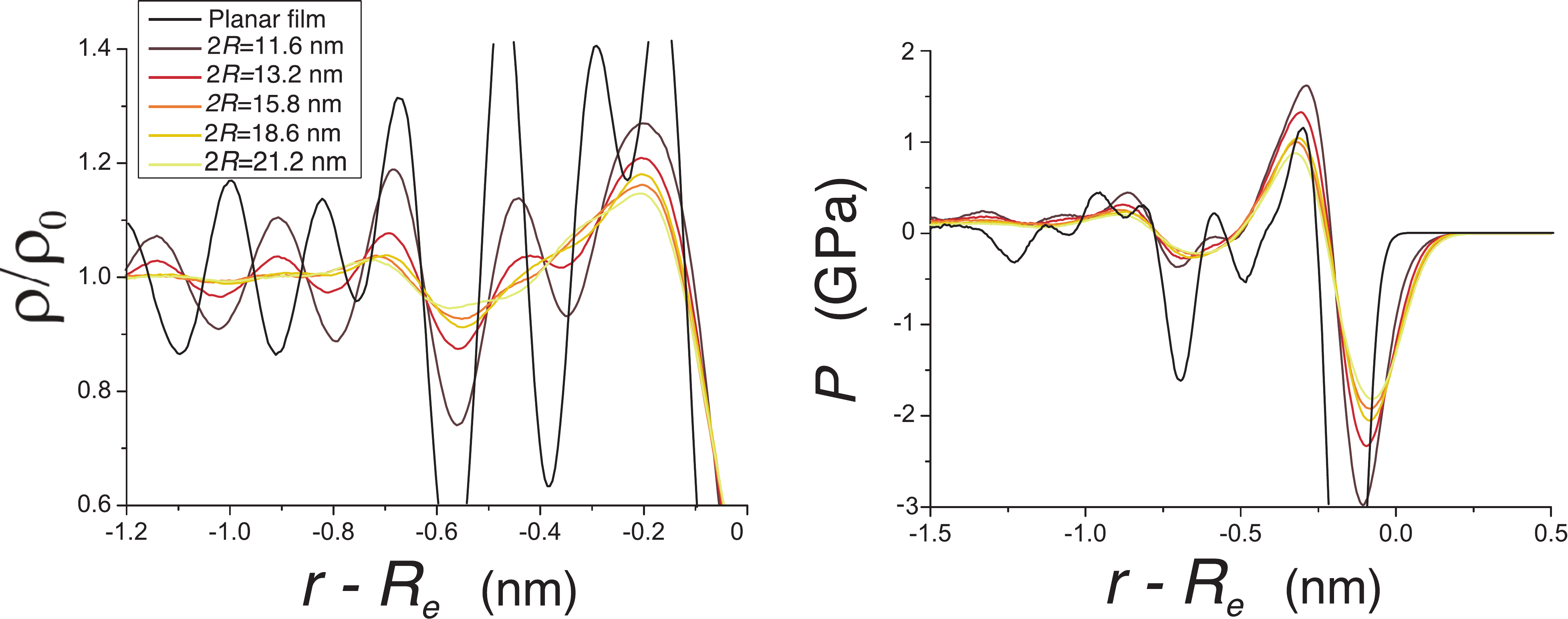}
 \caption{Same as in Fig.~\ref{fig:sizeVarDensityPressure873} but for $X_{Si}=33$\% and at $T=600\,$K. The combination of composition and temperature is such that the surface layers of both the liquid slab and the droplet are crystallized. \label{fig:sizeVarDensityPressure600}}
\end{figure*}

Figure~\ref{fig:sizeVarDensityPressure600} shows the size dependence of the density profile and the pressure distribution within the droplet below the surface crystallization temperature, $T=600$\,K and  $X_{Si}=33\%$. As in the thin films, the crystallization at the surface is confined to a single layer of thickness $0.32$\,nm (Fig.~\ref{fig:AuOrdering}). However, we see significant changes in the density profiles compared to the planar film, also plotted for comparison. The film is distinctly stratified over a thickness of 1.3\,nm, as shown earlier in Fig.~\ref{fig:densityFilms}a and \ref{fig:densityFilms600-700}, and this is also reflected in the pressure variation away from the surface layer. Within the droplets, the stratification is reduced compared to the films. We again observe an interplay between the surface segregation and the Laplace pressure at these sizes, and this is evident in the density profiles for the droplet with size $2R=21.2$\,nm.  While the extrema in the compression-tension at the surface associated with Au and and Si segregation are qualitatively different as the surface curvature modifies lateral order in the crystallized surface layer, for a given size the extent of the compression-tension extrema at the surface is enhanced relative to that observed above the surface crystallization temperature (Fig.~\ref{fig:sizeVarDensityPressure873}). The layer is strained to accommodate the curvature and its results in the formation of geometrically necessary extended defects such as grain boundary scars, evident in Fig.~\ref{fig:AuOrderingDroplet}. 

The Laplace pressure within the droplet leads to activation volume effects that enhance the stratification aided by surface curvature-induced strain in the crystalline surface phase. As the size is reduced, the crystallized layer is increasingly strained, the Laplace pressure within the droplet increases, and the linear density of the grain boundary scars on the surface increases. These three effects together set the trends in the surface segregation and pressure variation. The bulk pressure enhances the segregation due to the activation volume changes between the bulk and solid-like strained surface layer - the intensities associated with the monolayer Au and the subsurface Si increase, and we observe increased through-thickness stratification. The strain in the crystallized layer results in increase in the extrema of the compression-tension couple at the surface. As mentioned earlier, the increase is much more than that observed at $T=873$\,K due to the lateral order in the surface layer and this further enhances the effect of the activation volume changes between the surface and the bulk of the droplet. The grain boundary scars also serve as segregation sites for both Au and Si (Fig.~\ref{fig:AuOrderingDroplet}), and as their surface density increases at these small sizes, this has an added effect in amplifying the surface segregation. Overall, the combination of these effects increases overall surface segregation and stratification with decreasing size. At $2R=11.6$\,nm, the width of the stratified layer is comparable to that in the planar film, indicating that the strain in the surface layer together with the Laplace pressure effects play a dominant role in enhancing the surface crystalline order at small sizes. 

The combination of these effects at $T=600$\,K  also drives the variation in the average pressure within the droplet with surface curvature (not shown). Unlike the behavior at $T=873$\,K, the behavior is non-linear at high surface curvatures. The plot of the surface tension with effective radius does not show any conclusive trends at smaller sizes and exhibits large fluctuations, similar to the behavior observed in liquid-vapor interfaces close to a phase transition~\cite{surf:Giessen:2009}. The faceting of the crystallized layer implies anisotropy in the effective surface tension that additionally changes the size dependence. Our results show that the crystallized layer fundamentally modifies behavior of the surface layer in that its thermodynamic properties correspond to a solid-like layer that is sensitive to the surface strain as well as the size and distribution of grain boundary scars that dominate the surface structure.

\subsection{Surface Phase Transition}
Annealing and controlled cooling simulations allow us to study the thermodynamics of the surface crystallization in detail. We focus on planar films and droplets with composition $X_{Si}=33$\%. As before, the planar film simulations are performed on liquid slabs with in-plane periodic boundary conditions. The surface crystallization is observed over the two free surfaces using MD simulations. Each slab is relaxed for $2$\,ns at $800$\,K until the surface density profiles converge, and the average potential energy is calculated over 1\,ns equilibrium simulations. In order to extract the temperature range associated with the crystallization transition and to minimize the effects of high cooling rates inherent in this approach, equilibrated configurations at $800$\,K are quasi-statically cooled in decrements of $10$\,K. At each intermediate temperature, the simulation cell is again relaxed until the surface density profiles converge (within a few nanoseconds) and the average potential energy is extracted over a $1$\,ns equilibrium simulation thereafter. 
\begin{figure*}[htb]
\includegraphics[width=1.8\columnwidth]{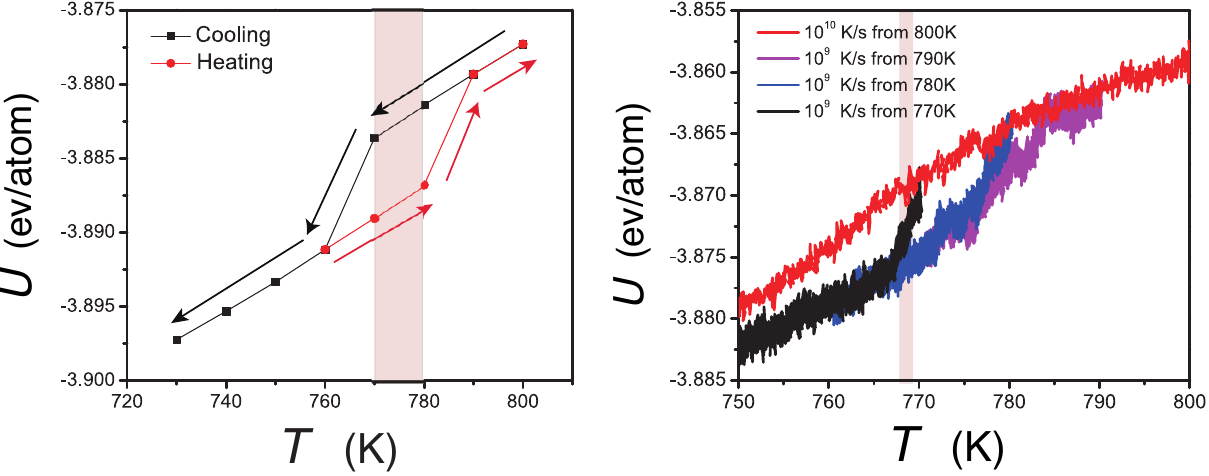}
\caption{(a) Potential energy per atom for a liquid AuSi ($X_{Si}=33$\%) slab cooled and then heated quasi-statically within the temperature range $T=730-800$K in $10$K intervals. At each intermediate temperature, the entire slab is relaxed within MD simulations for 2\,ns until the segregation profile in the vicinity of each of the two free surfaces reaches a steady-state. The transition temperature range lies in the range $T_s^\ast=760-790$\,K, is shown shaded in the figure. (b) Potential energy per atom for a $2R=11.6$\,nm liquid AuSi droplet of same composition as in (a), cooled with varying initial temperatures and quench rates. The transition temperature lies in the range $T_s^\ast=765-770$K.\label{fig:transitionFilm}}
\end{figure*}

Figure~\ref{fig:transitionFilm}a shows the variation in the average potential energy per atom versus temperature during the cooling cycle (black curve), averaged over the two free surfaces. The potential energy decreases with temperature, exhibits a sharp decrease from $T=770$\,K to $T=760$\,K that is correlated with formation of the crystalline surface layer, and then recovers the much slower decrease below $T=760$\,K. The transition temperature $T_s^\ast$ is strongly correlated with a slope change in $d\gamma_{lv}/dT$ (Fig.~\ref{fig:filmTension}).
The surface transition involves formation of 2D domains that rapidly grow and anneal into single crystalline phase at $T=760$\,K. The lattice parameter of the subsurface Si on the ordered phase is identical to that observed in equilibrium surfaces, $a=0.32$\,nm. The behavior is reminiscent of a first order transition associated with discontinuous change in the surface entropy at the transition temperature. The surface crystallization enthalpy per unit area is $0.19$\,J/m$^2$, averaged over the two surfaces. 

The evolution of the average potential as the liquid slab is heated back to $T=800$\,K is also shown in Fig.~\ref{fig:transitionFilm}a (red curve). The hysteresis in the average potential energy is negligible upto $T=760$\,K. We do not observe lateral melting of the crystallized layer beyond $T=760$\,K; rather the slab has to be superheated beyond $T=780$\,K before we recover the latent heat of formation of the surface layer. The melting is again a first order transition that occurs in the range $T=780-790$\,K. Comparison of the heating and cooling cycles show that the transition temperatures for formation and melting of the crystalline layer are separated by at least $\Delta T=10$\,K. The asymmetry also implies that the crystallization and melting of this layer is limited by nucleation of 2D islands.            

The droplet MD simulations are performed at fixed size and composition, $2R=11.6$\,nm and  $X_{Si}=33$\,\%. Starting with an equilibrated configuration at $T=800$\,K, the droplet is continuously cooled to $T=740$\,K and then reheated. The cooling cycle results are summarized in Fig.~\ref{fig:transitionFilm}b. At high cooling rates of the order of 10$^{10}$\,K/s, the potential energy decreases continuously with a change in slope below $T=780$\,K implying that the surface crystallization is no longer a first order transition. The cooling rates are admittedly high and can introduce artifacts. To this end, the MD simulations are performed at lower cooling rates. We observe a monotonic increase in the change of slope below $T=780$\,K. Quasi-statically decreasing in the initial temperature for the MD simulations to closer to $T=780$\,K further increases the slope change within a narrower temperature range. Fig.~\ref{fig:transitionFilm}b shows the thermal evolution of the average potential energy for droplets equilibrated at $T=790$, $780$ and $770$\,K, and at an order of magnitude lower cooling rate of 10$^{9}$\,K/s. In each case, the initial surface is laterally disordered before the droplet is quenched. For the droplet with initial temperature $T=770$\,K, lowering the temperature results in a sharp transition about $T_s^\ast=767$\,K that is consistent with a first order transition, and the recovery of a slower decrease in the potential energy after the transition is complete. The potential energy change associated with the surface crystallization is $0.15$\,J/m$^2$, a small decrease over that for the planar films. This follows from the fact that the crystalline droplet surface consists of multiple ordered domains separated by grain boundary scars that remain stable as the temperature is lowered below the transition temperature. Comparison with the planar film results reveals that the transition temperature is not sensitive to the surface curvature. Reheating the droplet to above the transition temperature shows that the superheating requited to melt the crystalline layer is smaller than that in the planar films (not shown), indicating that extended defects such as grain boundary scars lower the nucleation barrier for the order-disorder transition, as expected. 

\section{Discussion and Conclusions}
Premelting at crystalline interfaces such as crystal surfaces and interphase/grain boundaries is a fairly common phenomenon, dating back to observations of stable thin films of water on ice below the melting point~\cite{premelt:DahmenJohnson:2004}. The broken symmetry at these planar defects alters the thermodynamic state of the interfacial atoms and results in enhanced disordering of the interfacial region below the bulk melting point. Thermodynamic models on the stability of the wetting liquid layer are based on the competition between bulk and interface effects that account for the structure-based forces across the layer. The latter lead to a width dependent interface energy of the premelted layer, or a disjoining potential that sets its width~\cite{premelt:FensinHoyt:2010, fec:HoytUpmanyu:2010}. Analogously, prefreezing at solid-liquid interfaces is a reflection of the reduction in the disorder in the abutting liquid layers as the crystalline solid imposes its structure at the interface~\cite{sold:ReichertReiter:2000, tsf:Dijkstra:2004, tsf:LairdDavidchack:2007, sold:LohmanaThurnAlbrecht:2014}.

Prefreezing on liquid surfaces tends to be system specific. It has been observed in long chain liquid polymer films, and the behavior stems from surface-mediated anisotropic chain fluctuations and inter-chain interactions that facilitate crystallization on the surfaces~\cite{tsf:OckoSirotaKim:1993, sold:PrasadDhinojwala:2008}. Surface crystallization in binary metals is driven primarily by surface segregation of the species with the lower surface tension, resulting in an almost pure crystalline surface layer that serves as a precursor for bulk crystallization~\cite{tsf:YangRice:2007}. In the AuSi system, Si has a lower surface tension and that drives its segregation to the surface. However, pure Si crystallization into a covalently bonded silicene-like surface layer is energetically unfavorable as it requires a stabilizing crystalline substrate. Its formation is additionally subverted by the presence of Au, evidenced by the negative enthalpy of mixing, low eutectic temperature and glass forming properties of AuSi alloys~\cite{bmg:KlementWillensDuwez:1960}. As argued by Shpyrko et al.~\cite{tsf:ShpyrkoPershan:2007}, this is a reflection of the structural frustration in the bulk liquid which is partially released {\it via} surface crystallization. 

Above the surface crystallization transition temperature, our computations show that the surface layer undergoes stratification into alternating Si- and Au-rich layers. This appears to be the preferred mode for release of the structural frustration and it comes at the expense of lateral order along the surface. The spectrum of surface capillary fluctuations is consistent with the behavior of a liquid surface, that is $\langle|A(k)|^2\rangle \propto k^{-2}$. Evidently, the lateral disorder suppresses the anisotropy in surface energetics  resulting in liquid-like surface fluctuations. As the temperature is lowered below the crystallization transition temperature, the presence of the laterally ordered surface layer can lead to some deviations as the surface energy of fully crystalline interfaces typically varies with the surface orientation, and can also be modified by surface stresses~\cite{surf:Sander:2003, nano:SchmidNorskov:1995}. The anisotropy likely has a lesser impact for monolayer thick crystalline order on flat films, although
we do not observe statistically meaningful surface fluctuations over the tens of nanosecond time-scales accessible to the MD simulations, due to the combination of low temperature (and therefore diffusivities) and increased thickness of the crystallized layer. 

X-ray reflectivity studies by Shpyrko {\it et al.} on eutectic AuSi thin films have also reported the formation of a crystalline monolayer. The monolayer structure and stoichiometry are different (AuSi$_2$), yet the off-specular diffuse scattering is consistent with height-height correlations of a liquid-like surface layer~\cite{liquids:BraslauPershan:1988}. It follows then that the structure of the crystalline monolayer is able to absorb changes in its inclination in accordance with the underlying stratified liquid layers, and we therefore expect to observe similar liquid-like behavior in our computations at longer time-scales.

The computed values of surface tension are consistently lower than those reported in past experiments. Curve fits to the off-specular diffuse scattering data from X-ray reflectivity studies on eutectic thin films have yielded a value of $\gamma_{lv}=0.78$\,J/m$^2$.  Naidich {\it et al.} have employed the large drop (LD) method to measure the temperature and composition dependence~\cite{tsf:NaidichObushchak:1975}. They report a value of $\gamma_{lv}=0.86$\,J/m$^2$ at $X_{Si}=31\%$ and $T= 873\,K$, and a slight increase to $\gamma_{lv}\approx 0.9$\,J/m$^2$ closer to the eutectic composition $X_{Si}^E=21\%$. Note that these LD measurements are overestimates as they were made as the droplets were cooled to ambient temperatures, yet the discrepancy with the computed values is still large and cannot be accounted by the higher eutectic composition in the model AuSi system. Surface agents such as dissolved oxides and related contaminants have been implicated in the Si-rich surface structure, and they can certainly increase the surface tension~\cite{liquids:HalkaFreyland:2008}. However, we cannot disregard errors due to the empirical nature of the inter-atomic interactions used in the study. For example, AuSi model system is based on an EAM potential for pure Au that underestimates the liquid surface tension by 20\%~\cite{intpot:WebbGrest:1986}. We have addressed some of the shortcomings of the model potential in a previous study by including charge gradient corrections to the pure Au interactions~\cite{nw:WangUpmanyu:2013}. The changes in the normalized density profiles are negligible. Note that a similar disordered Au capping layer was observed in higher fidelity {\it ab initio} molecular dynamics studies on amorphous AuSi films~\cite{nw:LeeHwang:2010}, in accord with our computations. 

Nevertheless the computations capture key qualitative trends in the temperature and composition dependence of the surface tension of AuSi thin films that are in agreement with past experiments. The surface is Si-rich compared to the bulk composition over all temperature and composition ranges. The surface tension decreases with increases $X_{Si}$ - the trend was also observed in the LD studies and is again consistent with enhanced Si enrichment at the surface. At lower temperatures, the change in surface tension with temperature $d\gamma_{lv}/dT>0$ and serves as direct evidence of surface crystallization in these thin films. Immediately above the transition temperature, the film is significantly stratified and the slope $d\gamma_{lv}/dT$ decreases as the surface becomes less stratified, yet it is still positive and the slope changes its sign at much higher temperatures when the stratification is considerably suppressed. The extracted data on temperature dependence of the $\gamma_{lv}$ shows that segregation-induced stratification of liquid surfaces can be sufficient by itself for $d\gamma_{lv}/dT>0$  and therefore result in deviations from classical energetics of liquid surfaces.
\begin{figure*}[htp]
\includegraphics[width=1.5\columnwidth]{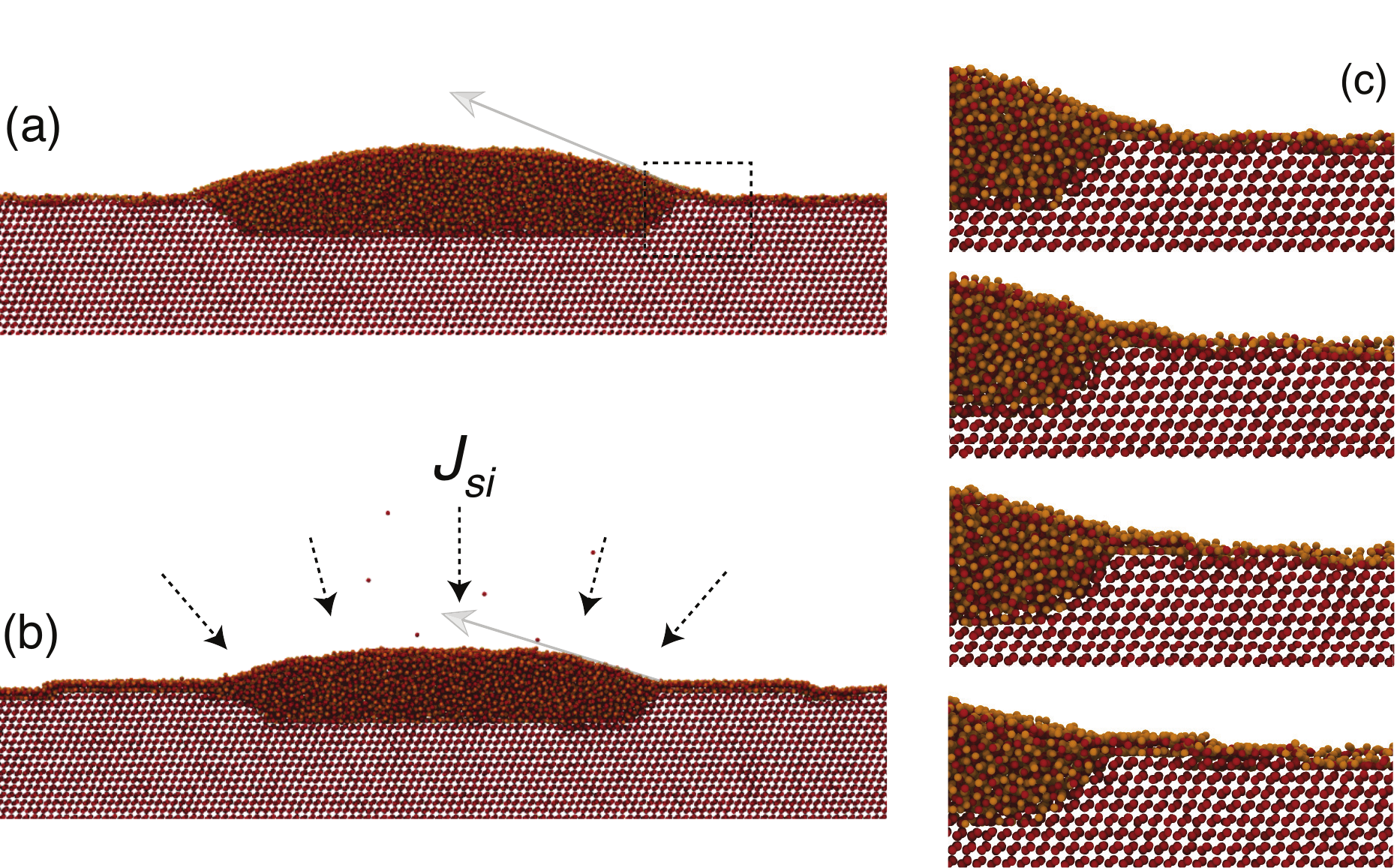}
 \caption{(a) Atomic configuration of a quasi-2D AuSi droplet at equilibrium on a Si(111) substrate at $T=873$\,K. The equilibrium composition of the droplet is $X_{Si}=39.5$\,\%. The solid-vapor layer is passivated by an Au layer. (b) Atomic configuration of the droplet-substrate system subject to constant surface current of $I_{Si}=2$\,atoms/ns at time $t=100$\,ns. Note the layer-by-layer growth at the solid-liquid interface, the reduction in the apparent contact angle, as indicated by arrows in (a) and (b), and partial growth of a new layer, referred to as a precursor wetting layer, at the solid-vapor interface. The droplet supersaturation varies between $X_{Si}=40-42$\,\%. (c) Temporal evolution of the precursor wetting layer following supersaturation during the first 20\,ns, showing the nucleation and growth of a precursor feet from the contact line that leads to the formation of a new layer.\label{fig:feetEvolution}}
\end{figure*}

The droplet studies reveal that above the surface crystallization transition, the Laplace pressure within the droplet modifies the segregation due to activation volume changes of both Au and Si between the bulk and surface, and we observe an overall increase in stratification at smaller sizes, i.e. the surface curvature modifies the activation volume differences between the bulk and the surface. Together with the fact that Si and Au segregation is associated with compressive and tensile stresses at the surface, this results in an overall increase in the through-thickness order with increasing surface curvature. The structural changes suggest that the surface tension is size dependent. The variation of the Laplace pressure differential across the surface layer shows that the surface tension varies non-linearly, yet the overall change is small. Extraction of the Tolman length, a measure of the size dependence, reveals that it is negative and of the order the interatomic distance. This length scale is the difference between the equimolar dividing surface  associated with density variations between the liquid-vapor (vacuum), and the mechanical surface through which the surface tension acts and that makes the Laplace equation exact~\cite{surf:Tolman:1948}. A negative Tolman length indicates the surface tension increases as the dividing surface moves outwards, towards the vapor phase. In effect, the surface segregation increases the effective radius of the droplet, $R_e-\delta$.

The surface curvature plays a more dominant role below the transition temperature. Unlike the thin films, the surface crystallized layer is strained and polycrystalline, consisting of grain boundary scars that become necessary to accommodate the large surface curvature. As a result, the order is considerably reduced compared to thin films and the effect cascades to a reduction in stratification. The compressive and tensile stresses associated with Si and Au segregation are qualitatively different due to the presence of surface strain and grain boundary scars. As the size is further reduced, we see a dramatic increase in stratification and the compression-tension couple at the surface due to an interplay between Laplace pressure driven surface segregation, surface strains that further amplify activation volume changes between the bulk and surface, and increasing density of grain boundary scars that serve as a sites for segregation of Au and possibly subsurface Si. The Laplace pressure variation with size becomes increasingly non-linear with size, indicating that the stratified layer with the surface crystalline phase no longer behaves as an effective liquid surface. Likely, the surface strain leads to surface stress based corrections to the surface energetics that render a simple capillarity based description inadequate.

Thermal cycling through the surface crystallization  transition {\it via} non-equilibrium MD simulation reveal that for both thin films and droplets, the laterally ordered surface layer crystallizes in the range $T=760-770$\,K. The surface crystallization and melting require undercooling and superheating respectively, indicating that the phase transition is mediated by the nucleation of 2D islands. In the case of droplets, the superheating required to melt the prefrozen surface is smaller, implicating the presence of grain boundary scars that serve as heterogeneous sites for melting of the crystalline layer and reduce the nucleation barrier. Although the thermal cycling rates are quite high in the MD simulations, the interaction potential energy change at the transition temperature becomes increasingly sharp as the cooling and heating rates are reduced, indicating that surface crystallization is a first order phase transition. The enthalpy change is not sensitive to the thermal cycling rates, and is $0.18$ and $0.15$\,Jm$^2$ for thin films and the $2R=11.6$\,nm  droplet respectively. We attribute the small decrease in the latent for the droplets to the presence of grain boundary scars.

The interplay between surface tension and composition of binary droplets is important for a range of applications. This is especially the case for vapor-liquid-solid growth of semiconducting nanowires. Although the size dependence of the surface tension is small for sizes of the AuSi nanoparticles studied here, it is sensitive to temperature and composition and the latter is especially important as the nucleation and growth of nanowires grown by these routes is mediated by supersaturated droplets whose composition oscillates during growth~\cite{nw:OhChisholmRuhle:2010, nw:SchwarzTersoff:2009, nw:WenTersoffRoss:2011}. 

To see if the effect is significant, we have have performed all-atom, quasi-2D MD simulations of the stability of an $2R=10$\,nm AuSi droplet on a Si(111) substrate. The simulations are performed above the surface crystallization temperature. The thickness of the quasi-2D droplet-substrate system is $3$\,nm and is sufficient to limit size effects associated with stratification of the surface layer. Figure~\ref{fig:feetEvolution}a shows the equilibrated AuSi droplet configuration. The droplet surface is decorated by a laterally disordered Au monolayer and stratified, consistent with density profiles in isolated droplets observed in our simulations. The substrate develops $\{113\}$ truncating facets as the droplet etches into the substrate until it reaches the equilibrium composition of $X_{Si}=39.5$\,\%. The presence of a monolayer thick wetting layer on the solid-vapor interface composed primarily of Au atoms is critical for the stability of the droplet. As validation, we have also performed simulations with pristine Si(111) solid-vapor surfaces. The droplet rapidly becomes Si-rich and serves as a source for the Au-rich wetting layer. It quickly destabilizes the droplet, suggesting that the equilibration of the Au surface chemical potential between the liquid-vapor and solid-vapor surfaces is a crucial ingredient for the stability of the droplet. Both the faceted morphology of Si substrates in equilibrium with AuSi droplet, and the stability of Au monolayer on the Si solid-vapor surfaces is consistent with past experiments and simulations~\cite{nw:FerralisMaboudian:2008, nw:HannonTromp:2006, nw:WangUpmanyu:2013}.  

The droplet is supersaturated by exposing it to a surface flux corresponding to a surface current of $I_{Si}=2$\,atoms/ns. The Si atoms are deposited at low kinetic energies ($1$\,eV/atom) and directly absorbed onto the droplet surface in MD simulations. Although the surface flux is high compared that due to much slower catalytic breakdown of Si precursors on AuSi surfaces~\cite{nw:KimRoss:2009}, it is low enough such that we observe layer-by-layer growth of the Si(111) substrate, consistent with the crystallization dynamics observed during VLS growth of Si nanowires. Figure~\ref{fig:feetEvolution}b shows the droplet-substrate system at $t=20$\,ns. The droplet is Si-rich with a silicon concentration $X_{Si}=42$\,\%. Step flow is evident at the Si(111) main facet in contact with the droplet leading to layer-by-layer growth of the substrate. 

The apparent contact angle of the droplet $\theta$ decreases (indicated by arrows), consistent with decrease in droplet surface tension with increasing $X_{Si}$. Following Young's balance along the horizontal, the contact angle is 
\begin{align}
\cos\theta = \frac{1}{\gamma_{lv}} \left(\gamma_{sv} - \gamma_{sl} \cos\alpha\right),
\end{align}
where $\alpha$ is the fixed angle between the $(111)$ and $(113)$  facets~\footnote{We have ignored torque terms associated with solid-liquid and solid-vapor interfaces in the balance under the assumption that the cusps in the $\gamma$-plot are sufficiently deep.}. Denoting $S$  as the spreading coefficient of the droplet, defined as 
\begin{align}
S = \gamma_{sv} - \gamma_{sl} \cos\alpha - \gamma_{lv}
\end{align}
the balance can be expressed as, 
\begin{align}
\cos\theta = 1 + \frac{S}{\gamma_{lv}}.
\end{align}
At equilibrium $S<0$ for the partial wetting droplet. 

The transient interval wherein the droplet builds up supersaturation and the Si growth is limited by nucleation, the decrease in surface tension with increasing $X_{Si}$ requires a smaller contact angle and also decreases the spreading coefficient $S$. The droplet is pinned by truncating facets and and volume conservation limits the extent of change in contact angle due to deviations from a spherical cap shape. Instead, the contact line adjusts by increasing the precursor wetting layer thickness locally, i.e. $\gamma_{sv}$ also decreases. The dynamics of the nucleation of this new layer from the contact line is shown in detail in Fig.~\ref{fig:feetEvolution}c. Evidently, the growth of this precursor feet is sustained by atomic surface diffusion of both Si and Au atoms from the droplet. It continues to grow to a finite distance before the Si supersaturation is absorbed by nucleation and step flow at the Si(111) solid-liquid interface. The precursor wetting layer serves as growth of crystallized layers that make up the base of the growing nanowire, suggesting that the eventual nanowire diameter is controlled by the extent of the precursor layer.  The growth of such microscopic precursor wetting layers is well known during non-equilibrium wetting of surfaces by non-volatile droplets~\cite{surf:PopescuCazabat:2012}. Our simulations highlight the role of the precursor layer, mediated by the compositional dependence of the droplet surface tension, and has ramifications for diameter selection of VLS grown nanowires. 

{\it Acknowledgements}: The authors are grateful for supercomputing resources available through Northeastern University and the Massachusetts  Green High Performance Computing Center (MGHPCC). The study was supported by a grant from National Science Foundation DMR CMMT Program (1106214). HW was also partially supported by the Thousand Young Talents Program of China.


\end{document}